\documentclass[12pt,twoside]{article}
\usepackage{amssymb,amsmath}
\usepackage{fleqn,espcrc1}

\usepackage{graphicx}
\usepackage[figuresright]{rotating}

\newcommand{\pdd}{P.~Danielewicz}
\newcommand{\etal}{{\em et~al.}}              %    et al.

\newcommand{\prl}[1]{Phys.~Rev.~Lett.~{#1}}
\newcommand{\npa}[1]{Nucl.~Phys.~{A#1}}
\newcommand{\prc}[1]{Phys.~Rev.~C {#1}}
\newcommand{\prd}[1]{Phys.~Rev.~D {#1}}
\newcommand{\zpa}[1]{Zeit.~f.~Phys.~{A#1}}

\newcommand{\beq}{\begin{equation}}
\newcommand{\eeq}{\end{equation}}
\newcommand{\bed}{\begin{equation}}
\newcommand{\eed}{\end{equation}}
\newcommand{\beqar}{\begin{eqnarray}}
\newcommand{\eeqar}{\end{eqnarray}}

\title{Flow and the equation of state of nuclear matter}
\author{\protect\pdd \address{
National Superconducting Cyclotron Laboratory and\\
Department of Physics and Astronomy, Michigan State University,
\\
East Lansing, Michigan 48824, USA}
        \thanks{e-mail:
danielewicz@nscl.msu.edu}}

\begin{document}

% typeset front matter
\maketitle

\begin{abstract}
The status of flow in heavy-ion collisions and of inference of
hadronic-matter properties is reviewed.
\end{abstract}

\section{INTRODUCTION}

Collective flow is a~motion
characterized by space-momentum correlations of dynamic origin.
It is of interest in collisions because it may
tell us about pressures generating that motion and about the
equation
of state (EOS) and other properties of the strongly-interacting
matter. The~flows that have been identified thus far are
radial,
sideward, and elliptic.

The role played by the pressure $p$ in the dynamics may be
inferred by
examining the relativistic Euler equation in the frame of the
matter, defined by vanishing collective velocity $v=0$:
\bed
(e + p) \, \frac{\partial}{\partial t} \, \vec{v} =
- \vec{\nabla} p \, .
\label{eq:newton}
\eed
Here $e$ is energy density.
This equation is an analog of the Newton equation
and enthalpy $w=e+p$ plays the role of mass.
Pressure plays the role of potential for the collective motion.

\begin{figure} %[htb]
\centerline{\includegraphics[angle=0,origin=c,
width=.70\linewidth]{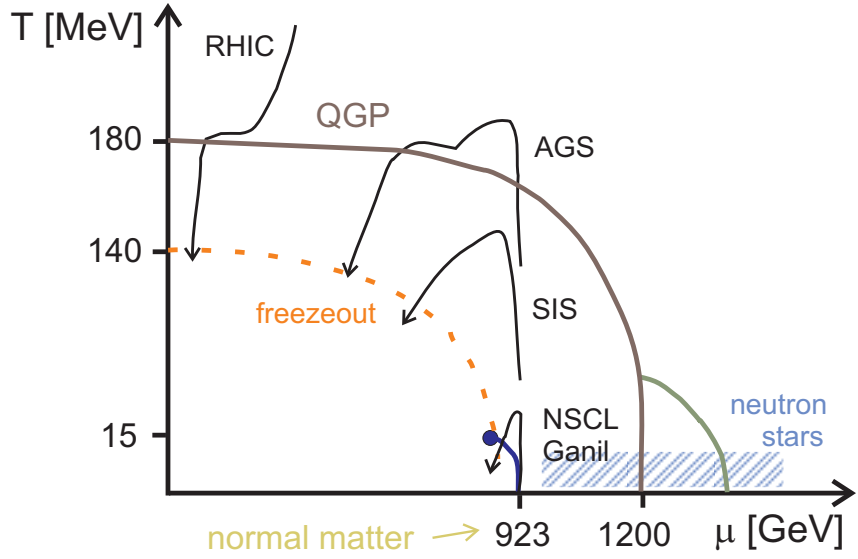}}
\caption{Temperatures $T$ and baryon chemical potentials $\mu$
explored in central heavy-ion reactions.}
\label{fig:muT}
\end{figure}

Recent interest in the flows and EOS is due to
the possibility of identifying the phase transition to
quark-gluon plasma.
Figure~\ref{fig:muT}
shows a schematic diagram of temperatures $T$ and baryon
chemical
potentials $\mu$ explored in central heavy-ion reactions with
the
indicated possible crossings of the transition at different
accelerators.

\begin{figure} %[htb]
\centerline{\includegraphics[angle=90,origin=c,height=3.0in,
width=.70\linewidth]{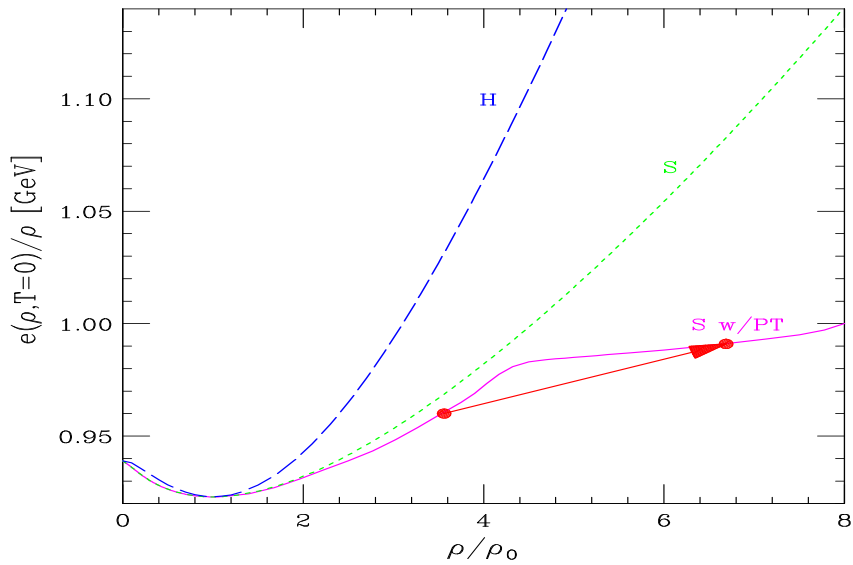}}
\caption{Energy per baryon $e/\rho$ as a~function of baryon
density~$\rho$, in units of normal density $\rho_0$,
for a~soft EOS (S), hard EOS~(H), and a~soft EOS with
a~high-density phase transition (S~w/PT).}
\label{fig:eosnn00}
\end{figure}

In terms of baryon density $\rho$, the pressure is:
\bed
p = \rho^2 \, \left. \frac{\partial (e/\rho)}{\partial
\rho} \right|_{s/\rho} \, ,
\eed
where $s$ is the entropy density.  Expected changes in the
energy per baryon $e/\rho$, if there were a phase transition
at $T=0$, are qualitatively indicated in
Fig.~\ref{fig:eosnn00}.
The arrow in the figure represents the Maxwell construction.

Within and
above any first-order phase-transition, or a~transitional
region, a~physical system exhibits a~softening, i.e.\ the
ratios
$p/e$ and $p/w$ drop.  This is due to the fact that the
pressure $p$ changes {\em continuously} across the transition,
while
$e$ jumps together with entropy $s$ and density~$\rho$.  For
example, in the baryonless matter we have:
\bed
dp = s \, dT \, , \hspace*{2em} de = T \, ds \, .
\eed
With an increase in temperature across a~transition,
the~entropy jumps and so does the energy.  However, for
the pressure only the derivative changes.  Analogous changes
across
a $T=0$ transition are deduced in Fig.~\ref{fig:eosnn00}.
Following~(\ref{eq:newton}), the signature of a
phase transition should be a slowed down pace of the
development of the nuclear collective motion within and above
the phase transition.

Equilibrium, to which EOS and our
considerations so far refer to, is never precisely reached in
reactions.  This creates both difficulties and opportunites.
On one hand, extrapolations must be done to the equilibrium
limit.  On the other hand, the flow may be exploited in
exploring the
transport properties of hadronic matter, related to in-medium
cross sections.
The flow may be studied to a variable degree of detail.
A~study of nucleon flow as a function of transverse momentum,
for example, may reveal the dependence of the mean field (MF)
felt by nucleons on momentum~\cite{ham90}.

We now turn to the different types of flow.

\section{RADIAL EXPANSION}

\begin{figure} %[htb]
{\includegraphics[angle=0,origin=c,
width=.40\linewidth]{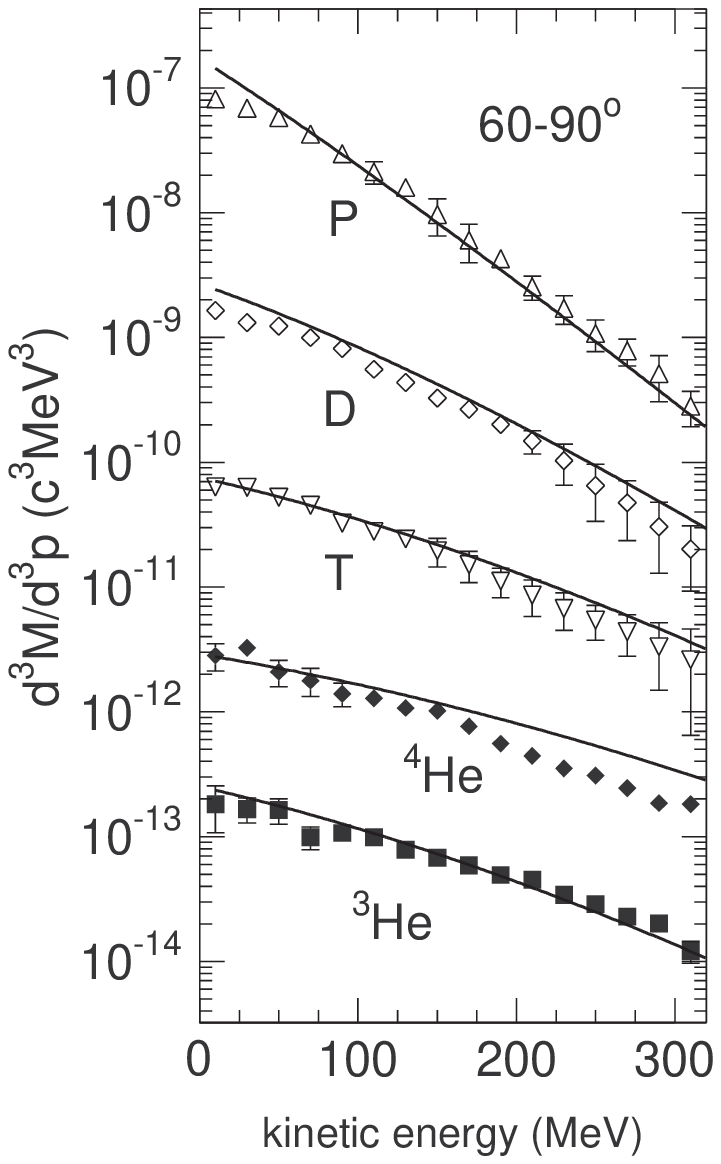}}\hspace*{.5em}
{\includegraphics[angle=0,origin=c,
width=.57\linewidth]{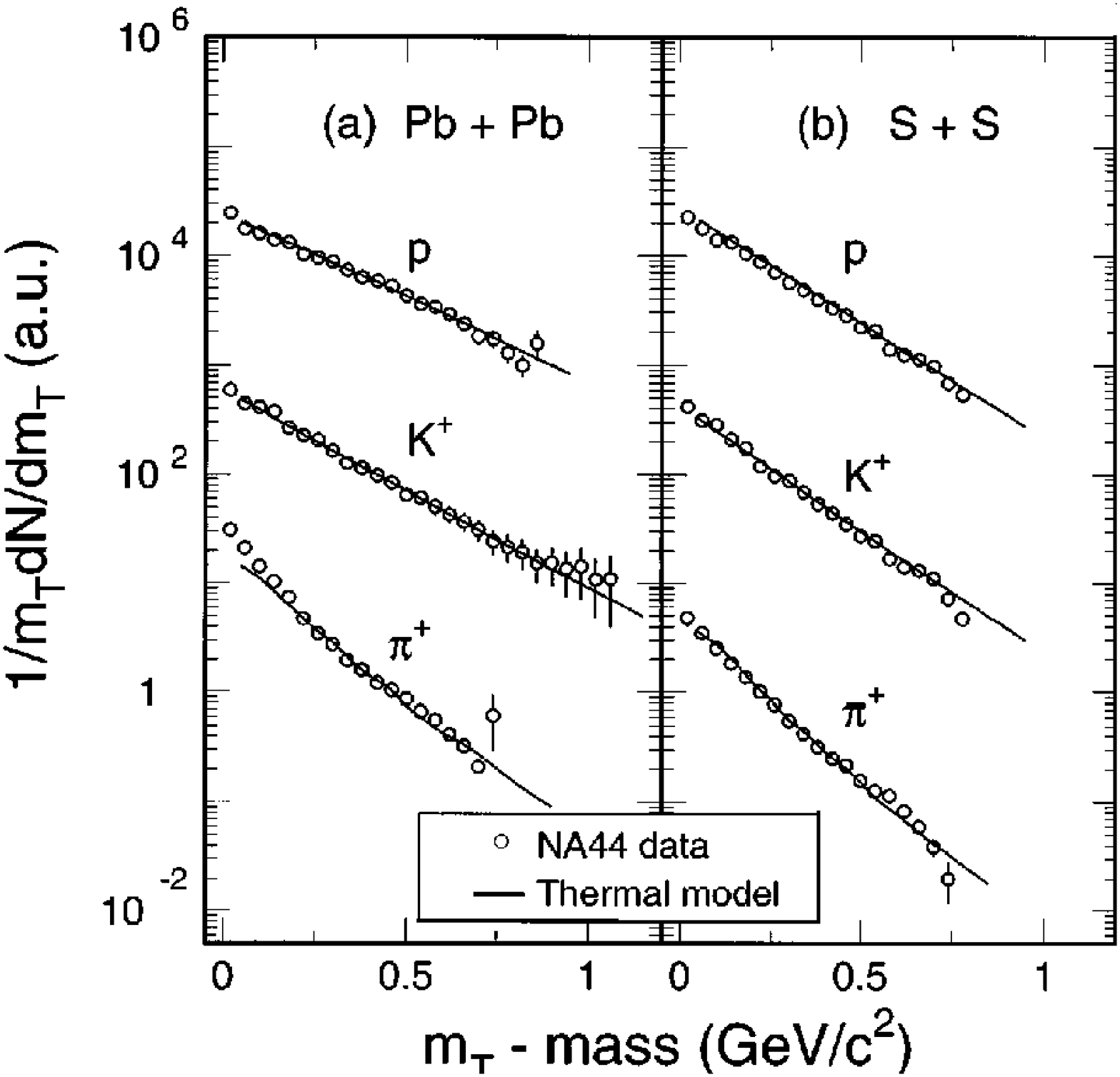}}
\caption{Transverse momentum distributions in central
collisions at 250~MeV/nucleon (left) and at 158~GeV/nucleon
(right) from the measurements of the FOPI~\protect\cite{pog95}
and of the NA44~\protect\cite{bea97} Collaborations,
respectively.}
\label{fig:pT}
\end{figure}

The collective radial expansion is often assessed by looking
for deviations of momentum distributions, especially
transverse, from thermal.  The~momentum distributions are
commonly described in terms of the simple Siemens-Rasmussen
\cite{sie79}
formula, or its derivatives,
\beq
\epsilon \, {dN \over d{\bf p}} \propto
{\rm e}^{-\gamma \epsilon /T} \left\lbrace{1 \over pv}
\left(\gamma
\epsilon + T \right) {\rm sinh} \, {\gamma p v \over T} - {\rm
cosh} {\gamma p v \over T} \right\rbrace \, .
\eeq
A~safer assessment of the radial flow is by comparing spectra,
or average energies,
of particles with different mass.   The~higher the mass,
the stronger is the effect of collective expansion and flatter
the
distribution (more spread-out by the collective velocity), and
higher
the average energy.  Transverse distributions flattening with
the particle mass have been seen in central collisions of heavy
nuclei at beam energies ranging from below 100~MeV/nucleon to
above 100~GeV/nucleon, cf.~Fig.~\ref{fig:pT}.
The~larger-mass distributions are sharper
in the lighter than in the
heavier system, at the high energy, indicating a~weaker
collective expansion in the lighter system.

\begin{figure}[htb]
\centerline{\includegraphics[angle=90,
width=.55\linewidth]{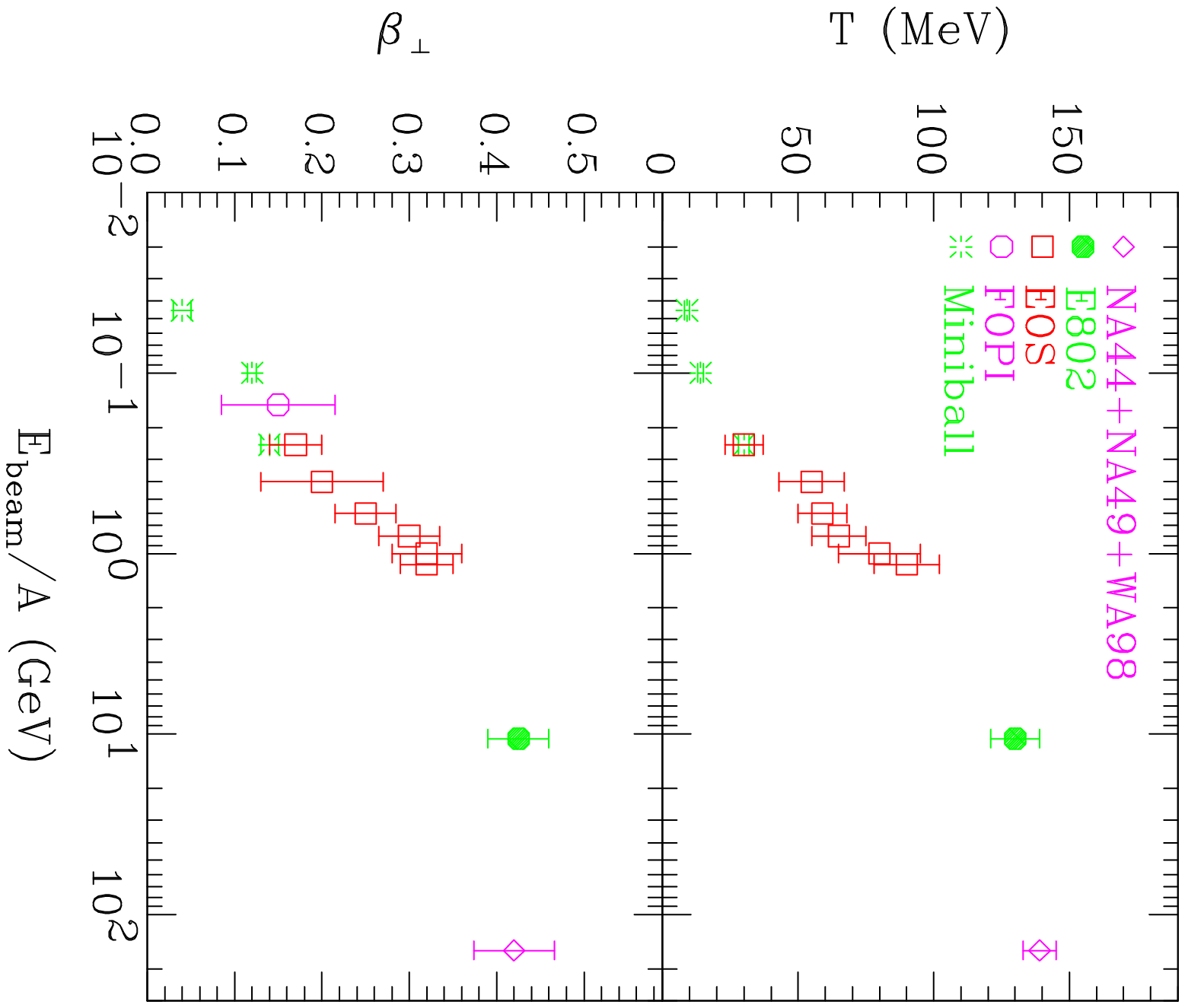}}
\caption{Excitation functions of transverse temperature (upper
panel) and transverse
collective velocity (lower panel) at midrapidity in heavy
systems.}
\label{fig:bT}
\end{figure}

Figure~\ref{fig:bT} displays excitation function of transverse
temperature and of velocity in heaviest systems.  The~velocity
saturates at AGS energies, possibly due to meson production and
progressing transparency.

Of interest is the possible use of the radial expansion in the
determination of EOS.  It must be remembered that the
separation into the collective and thermal energies occurs at
freeze-out, when collisions become infrequent.  Let us consider
first the situation at low energies.  If the EOS is soft and
pressure low, the expansion is slower than for a~stiff EOS, but
then one just needs to wait longer for same observable values
to
emerge at freeze-out.  To tell the difference, one needs some
timing device.
As such devices might serve the persistence of longitudinal
motion at
high energies or the early strange particle emission.  Till now
though, their exploitation in the EOS determination has been
limited~\cite{kol00}.

\section{SIDEWARD FLOW}

\begin{figure} %[htb]
\begin{minipage}[t]{50mm}
\centerline{\includegraphics[angle=0,
width=44mm]{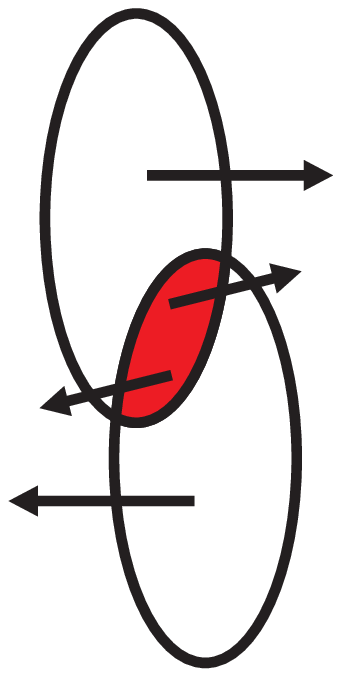}}
\caption{In-plane particle deflection.}
\label{sideflow}
\end{minipage}
\hspace{\fill}
\begin{minipage}[t]{105mm}
\centerline{\includegraphics[angle=0,
width=85mm]{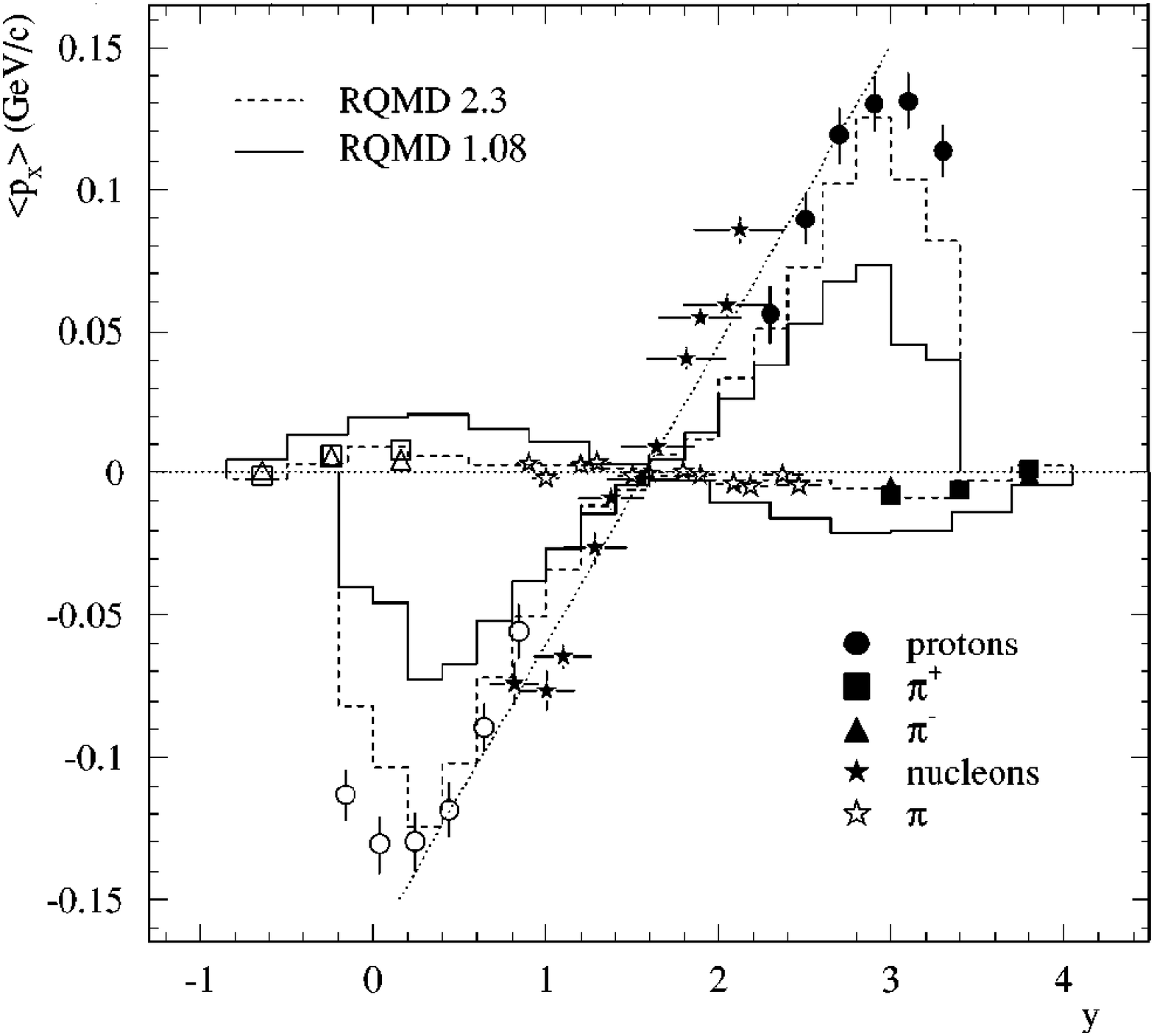}}
\caption{Average
in-plane transverse momentum component
as a function of rapidity in central Au + Au collisions at
10~GeV/nucleon~\protect\cite{bar97}.}
\label{FF}
\end{minipage}
\end{figure}

Sideward flow is a~deflection of forwards and backwards moving
particles, away from the beam axis, within the reaction plane.
The situation in reactions is schematically illustrated in
Fig.~\ref{sideflow}.
For the compressed and excited
matter in a~central region it is easier to
get out to the vacuum on one side of the beam axis than on the
other. Eagerness to get out will be enhanced by high generated
pressure but also by the momentum dependence
of MFs before
the equilibration takes place.
The~ability to get out depends on the inter-particle cross
sections.

The sideward flow is often represented in terms of the
mean in-plane component of transverse momentum at a~given
rapidity, $\langle p^x (y) \rangle$, and additionally
quantified in terms of the~derivative at the midrapidity, see
Fig.~\ref{FF}:
\beq
F_y = {d \langle p^x \rangle \over d y}
\hspace*{2em}\mbox{or}\hspace*{2em}
F = {d \langle p^x \rangle \over d (y/y_B)} \, .
\eeq
The normalization of the rapidity to the beam in the
derivative
enhances, somewhat artificially, the strength of dynamic
effects at high energies relative to low.

\begin{figure} %[htb]
\centerline{\includegraphics[angle=90,origin=c,
width=.6\linewidth]{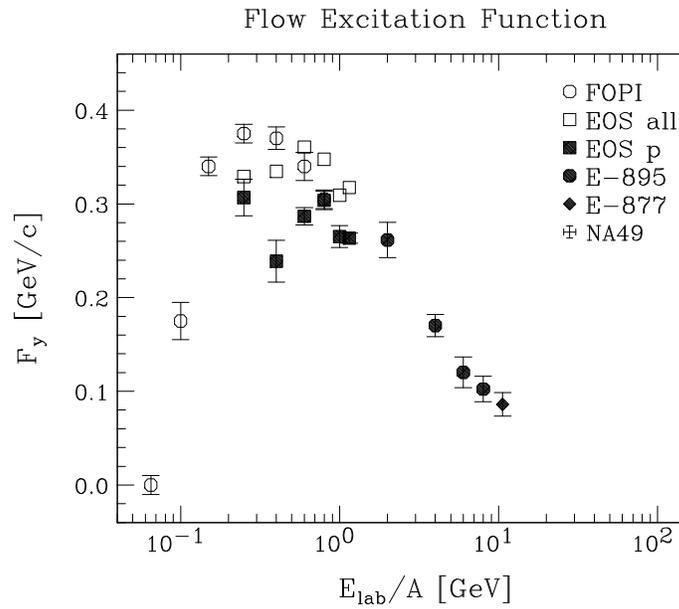}}
\vspace*{-.3in}
\caption{Excitation function of sideward flow in
central collisions of heavy nuclei.}
\label{fig:Fy}
\end{figure}

In transport models, it is directly observed that the
production of sideward flow is shifted towards the high density
phase~\cite{jan92} as compared to the radial flow~\cite{dan92}.
The~sideward flow thus has more potential in the EOS
determination than the radial flow.  The~flow excitation
function is represented in Fig.~\ref{fig:Fy} and the flow is
seen to be maximal between 0.1 and 10~GeV/nucleon.

\section{SECOND-ORDER OR ELLIPTIC FLOW}

\begin{figure}[htb]
\centerline{\includegraphics[angle=0,origin=c,
width=.70\linewidth]{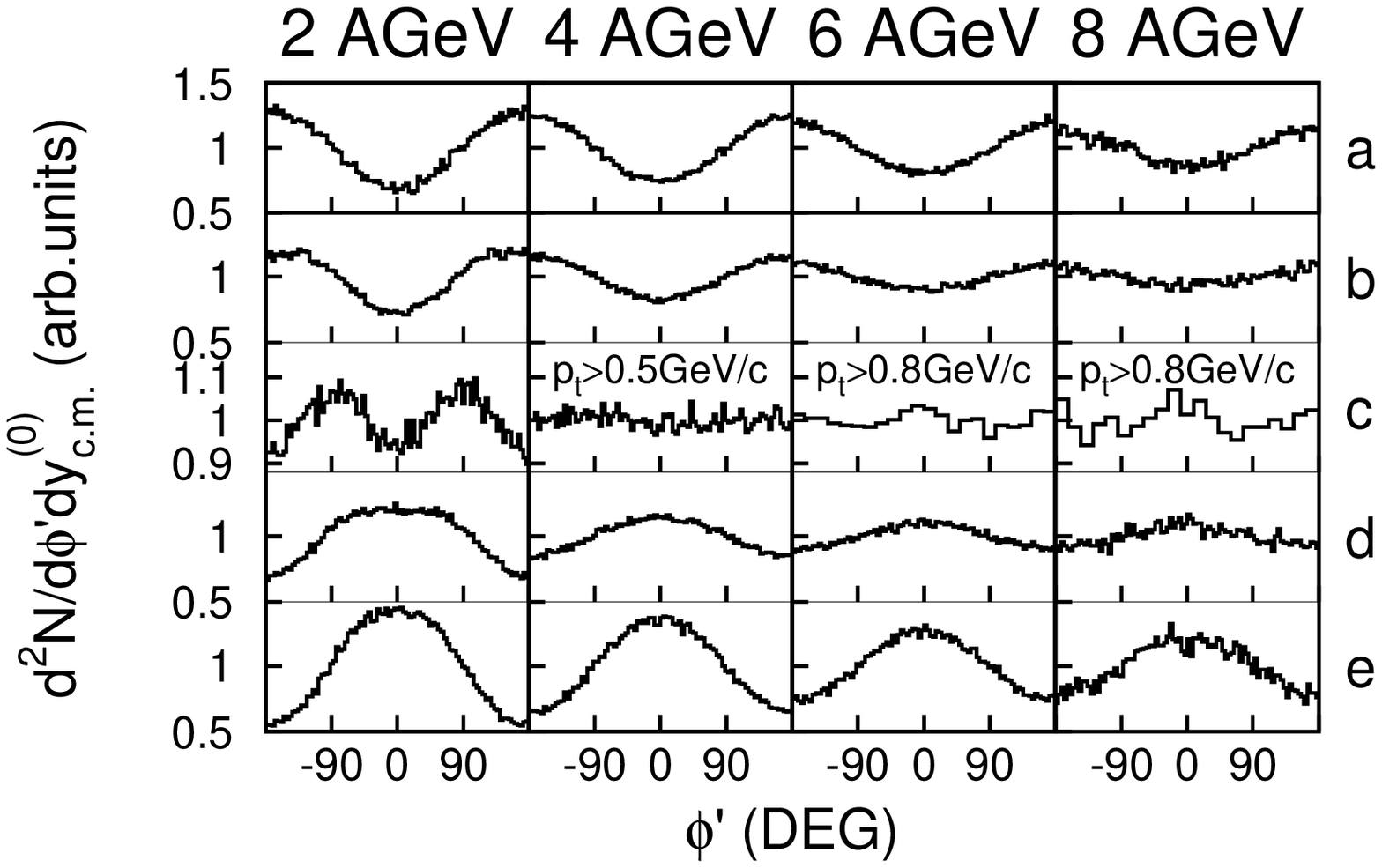}}
\caption{
Azimuthal distributions, with respect to the reconstructed
reaction plane, of protons emitted from semicentral Au + Au
collisions~\protect\cite{pin99}
in the rapidity intervals of $-0.7 < y/y_{Beam} < -0.5$ (a),
$-0.5 <
y/y_{Beam} < -0.3$ (b), $-0.1 < y/y_{Beam} < 0.1$ (c), $0.3 <
y/y_{Beam} < 0.5$ (d), and $0.5 < y/y_{Beam} < 0.7$ (e).}
\label{pink}
\end{figure}

The elliptic flow is typically studied at midrapidity and
quantified in terms of $v_2$:
\beq
v_2 = \langle \cos{2 \phi} \rangle \hspace*{3em}
\left(\, v_n = \langle \cos{ n \phi} \rangle \, \right)
\, ,
\eeq
where $\phi$ is the azimuthal angle relative to the reaction
plane.  The~second-order flow may offer a~better chance for the
EOS determination than the first-order sideward flow, because
it involves less of the uncertainties in the opposing streams
of matter moving past each other.  Typical azimuthal patterns
at midrapidity may be seen in Fig.~\ref{pink}.

\begin{figure}[htb]
\centerline{\includegraphics[angle=0,
width=.75\linewidth]{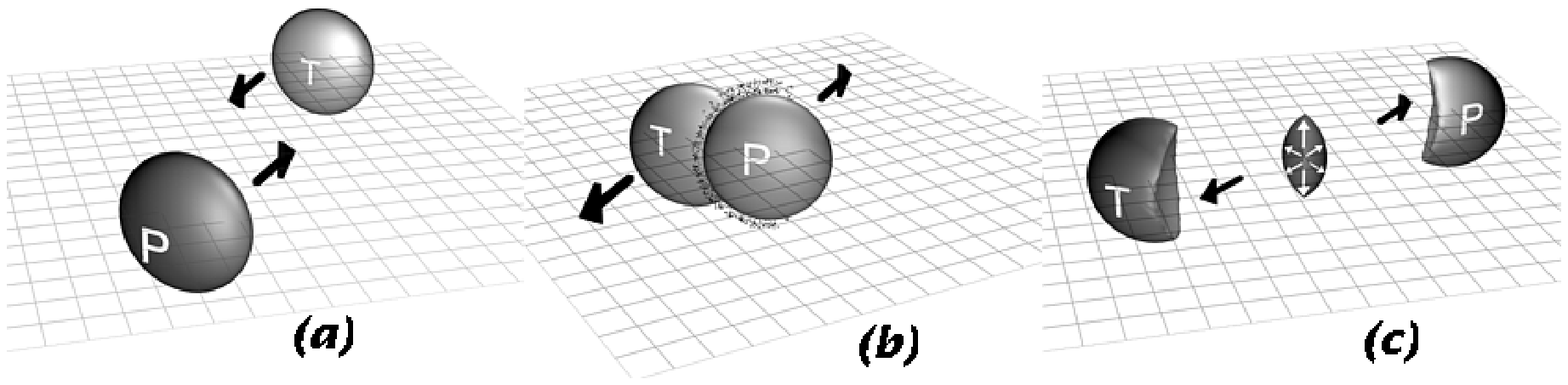}}
\caption{
Collision of two Au nuclei at relativistic energies.
Time shots are shown for an instant before the collision~(a),
early in the collision~(b), and late in the collision~(c).
}
\label{roy1}
\end{figure}

At AGS energies the elliptic flow results from a~competition
between the early squeeze-out when compressed matter tries to
move
out in the~direction perpendicular to the reaction plane and
the late-stage in-plane emission
associated with the shape of the participant
zone~\cite{oll92}, cf.~Fig.~\ref{roy1}.
The~squeeze-out contribution to the elliptic flow depends,
generally, on
the pressure $p$ built-up early on, compared to the energy
density $e$, see~(\ref{eq:newton}), and on the passage time
for the
spectators.
When the heated matter is exposed to the vacuum in
the transverse direction, a~rarefaction wave moves in putting
the
matter into motion.
The~time for developing the expansion is
then $R/c_s$, where $R$ is the nuclear radius and $c_s =
\sqrt{(\partial p / \partial e)_{s/\rho}}$ is the speed of
sound.
The~passage time for spectators, on the other hand, is of the
order of $2R/(\gamma_0 \, v_0)$, where $v_0$ is the spectator
c.m.\ velocity.  The~squeeze-out contribution to the elliptic
flow should then reflect the time ratio
\beq
{c_s \over \gamma_0 \, v_0} \, .
\label{ratio}
\eeq

\begin{figure}[htb]
\centerline{\includegraphics[angle=90,origin=c,height=3.5in,
width=.65\linewidth]{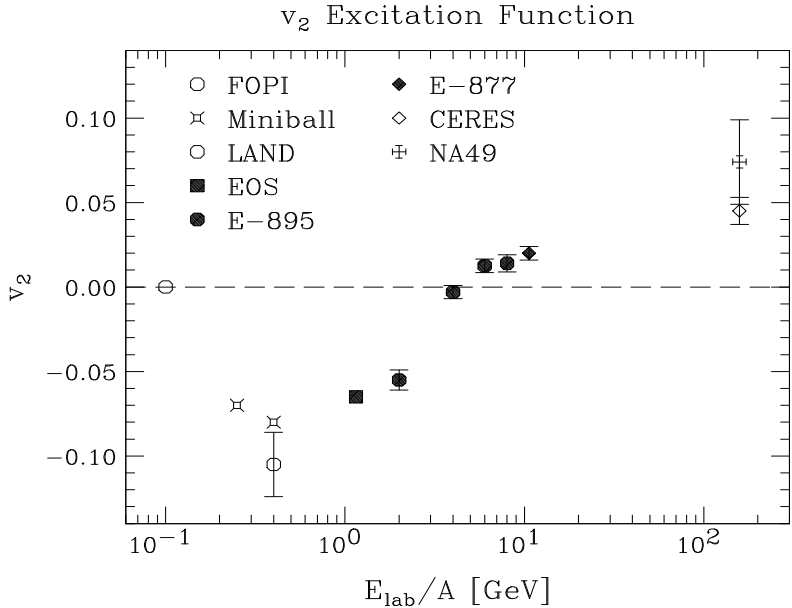}}
\vspace*{-.3in}
\caption{Elliptic-flow excitation function.}
\label{v2}
\end{figure}

\begin{figure}[htb]
\centerline{\includegraphics[angle=0,origin=c,height=4.5in,
width=.70\linewidth]{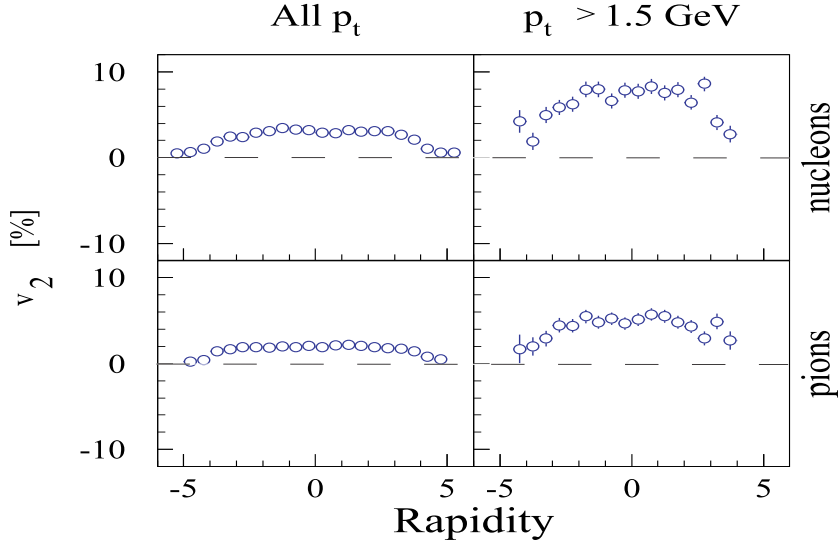}}
\vspace*{-.1in}
\caption{Elliptic-flow coefficient for nucleons and pions from
RQMD simulations of $b= (5-10)$~fm Au + Au collisions at
$\sqrt{s}$ = 200~GeV \protect\cite{sne99}.}
\label{fig:v2RHIC}
\end{figure}

The result~(\ref{ratio}) gives hope that significant changes in
the dependence of pressure on energy density, such as
associated with a~phase transition,\footnote{Discussion of the
physics of refraction into mixed phase or across the transition
may be found in~\protect\cite{dan87}.} might be revealed in the
variation of the elliptic flow.  Overall, the
squeeze-out
contribution should decrease as a~function of energy with the
flow becoming positive, $v_2 > 0$.  The elliptic-flow
excitation function is displayed in Fig.~\ref{v2}.
It is seen that, indeed, while dominated by squeeze-out at
moderate energies, the elliptic flow becomes positive at high
energies.  Whether or not any changes in $v_2$ with energy
might
be associated with any phase transition requires comparisons to
transport-model calculations to assess the magnitude of
such possible changes.  It should be noted, due to the
expansion of elliptically shaped participant zone exposed from
all sides to the vacuum, Fig.~\ref{roy1}(c), the~in-plane
elliptic flow is expected to be present at RHIC energies, see
Fig.~\ref{fig:v2RHIC}.

To assess the practical quantitative level of sensitivity of
flow
observables to EOS and to other characteristics of the medium,
we now turn to transport model simulations.

\section{TRANSPORT-MODEL COMPARISONS}

We shall examine results of a model formulated within
relativistic Landau theory, with nucleon, pion, delta and $N^*$
degrees of freedom.  The~Wigner distribution functions $f$ of
those particles follow a~set of Boltzmann equations, of the
same general for relativistically, as nonrelativistically,
\beq
{\partial f \over \partial t} + {\partial \epsilon_{\bf p} \over
\partial {\bf p} }
\, {\partial f \over \partial {\bf r}} -
{\partial \epsilon_{\bf p} \over
\partial {\bf r} } \, {\partial f \over \partial {\bf p}}
  =  I \,  .
\label{BE}
\eeq
The single particle energies~$\epsilon$ are functional
derivatives of an energy functional specified in terms of the
Wigner functions~$f$.  Specific parametrizations of the
functional lead to different EOS, such as in
Fig.~\ref{fig:eosnn00}, and different optical potentials
\beq
\label{Uopt}
U^{opt}(p) = \epsilon(p) - m - T(p) \, ,
\eeq
where $T$ is kinetic energy, that either depend or do not on
the momentum~$p$ in the medium frame.

\begin{figure}[htb]
\centerline{\includegraphics[angle=90,
width=1\linewidth]{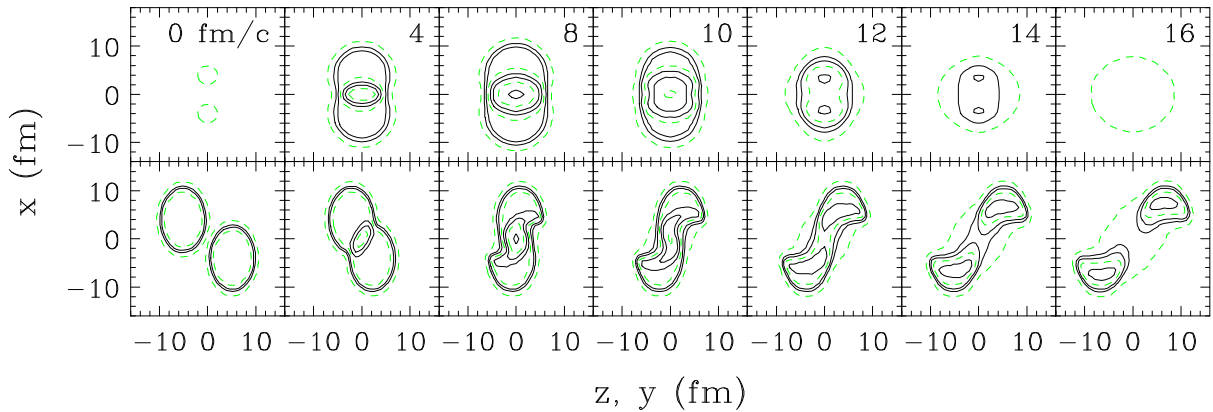}}
\caption{
Contour plots of baryon density in the 2 GeV/nucleon Au + Au
reaction at $b=8$~fm in the reaction plane (bottom panels)
and in the plane through the system center orthogonal to the
beam axis (top panels).  The~simulation was done assuming a
hard EOS and momentum-dependent $U^{opt}$.
The~density contours are shown at $\rho/\rho_0 = 0.1$, 0.5, 0.8
and then at 0.4 intervals.
Every third contour, i.e.\ for $\rho/\rho_0 =0.1$, 1.2,
and~2.4, is represented by a~dashed line.  The~remaining
contours are represented by solid lines.
Numbers in the figure indicate time in~fm/c.
}
\label{fig:con}
\end{figure}

\begin{figure}%[htb]
\begin{minipage}[t]{71mm}
\centerline{\includegraphics[angle=90,height=65mm,
width=70mm]{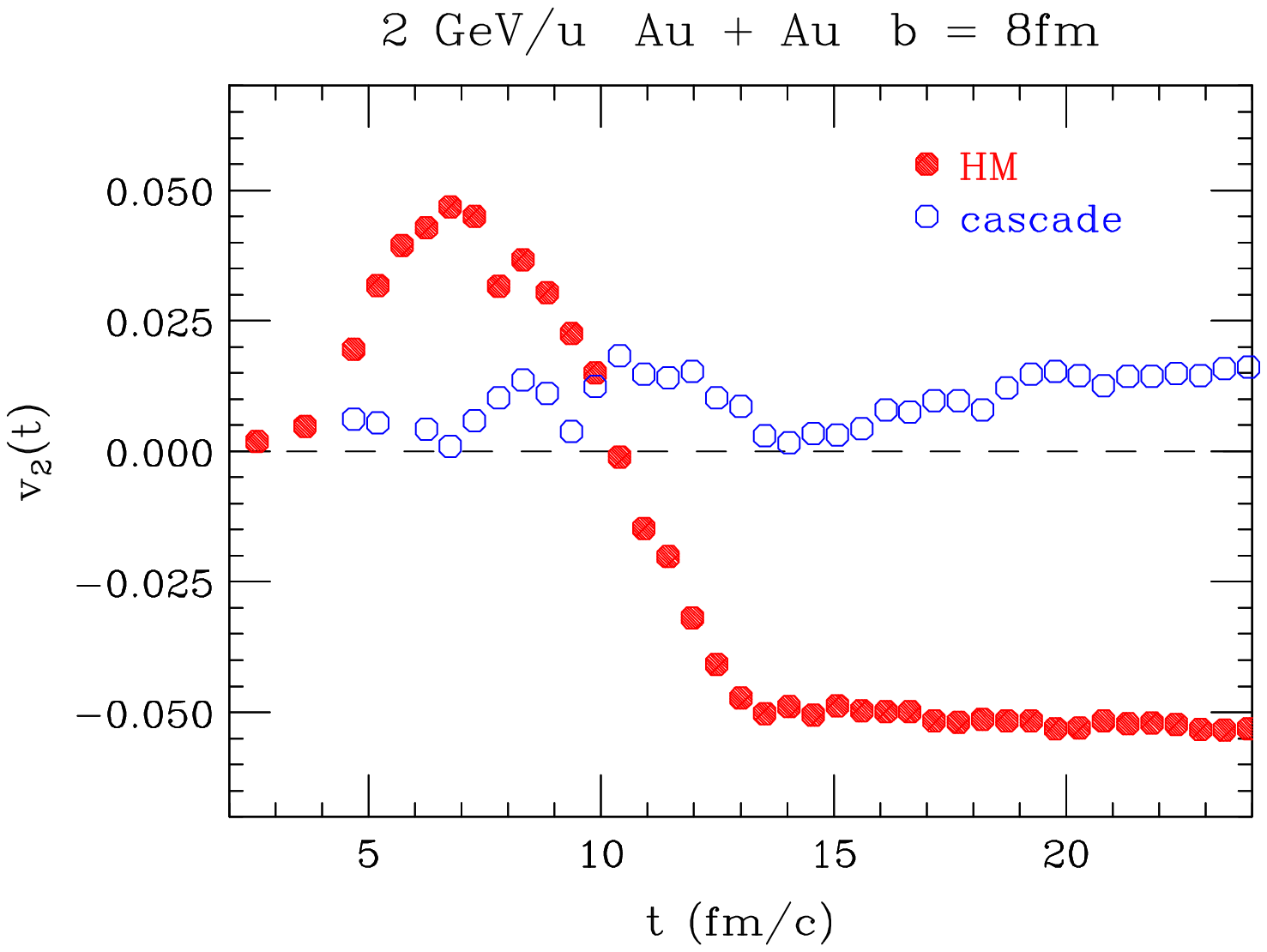}}
\caption{Temporal evolution of the midrapidity elliptic
anisotropy in 2 GeV/nucleon Au + Au collision at $b=8$~fm
in the cascade mode of the model (open circles) and for
a hard EOS with momentum dependence in the MFs
(filled circles).}
\label{fig:v2ti}
\end{minipage}
\hspace{\fill}
\begin{minipage}[t]{83mm}
\centerline{\includegraphics[angle=0,height=71mm,
width=82mm]
{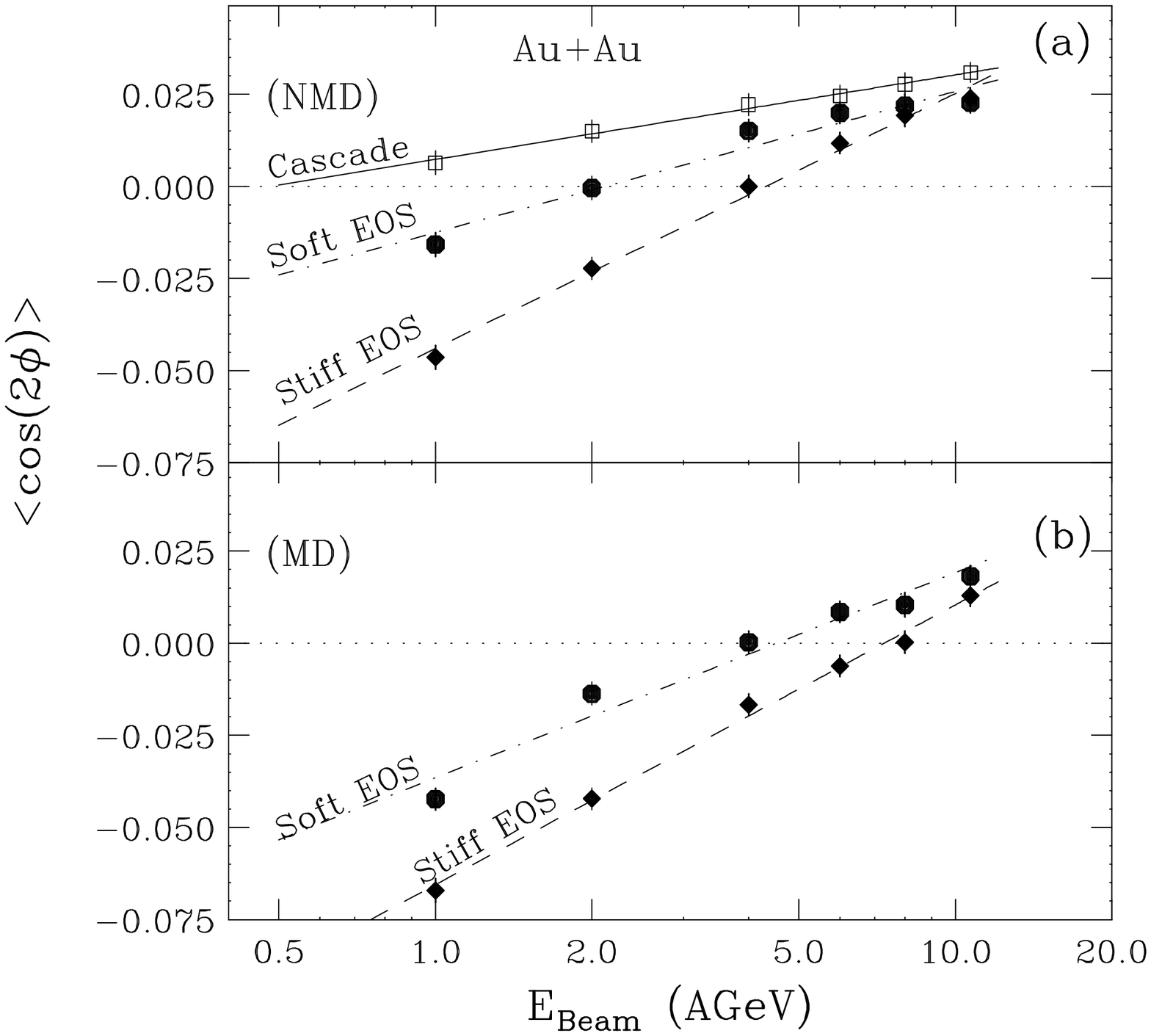}}
\caption{
           Calculated elliptic flow excitation functions for Au +
Au reactions. Panels~(a) and~(b) show, respectively, the~functions obtained
without~(NMD) and with~(MD) the momentum dependent forces.
The~filled circles, filled diamonds, and open squares indicate,
respectively, results obtained using a~soft EOS, a~stiff EOS,
and by neglecting the MF.  The~straight lines show logarithmic
fits.
}
\label{fig:fig3}
\end{minipage}
\end{figure}

Typical single-particle observables from the
model~\cite{dan99},
for different EOS, such as transverse momentum or rapidity
distributions, compare favorably to data~\cite{ahl98}.
Figure \ref{fig:con}, with baryon density contour
plots from a~simulation of the 2~GeV/nucleon Au + Au reaction
at $b=8$~fm, helps to understand the
generation of elliptic flow.  The~simulation was done assuming
a~hard EOS and a~momentum dependence in the baryonic MFs.
The~participant matter gets compressed in
Fig.~\ref{fig:con} to $\lesssim 3\, \rho/\rho_0$ and then
explodes when spectator pieces are still present.  When the
simulation is done in the cascade mode of the model,
the~participant matter gets compressed to higher densities and
it decompresses in quite a sluggish manner.  At the completion
of the decompression the spectator pieces are far gone.
Figure \ref{fig:v2ti}
shows next the evolution of the midrapidity elliptic anisotropy
in the two simulations.  In either of the simulations,
the~anisotropy begins to grow in the in-plane direction.  In
the hard-EOS calculation, the growth is rapid and then the
anisotropy changes its sign.  In the cascade calculation the
growth is slow and it continues till late times.
These findings may be understood as follows.
The~expansion of the participant region first grows
primarily in the in-plane direction as in
that direction the pressure gradients are the
largest~\cite{oll92}.  If the expansion is rapid, though, the
matter encounters the spectator pieces on its way.
Then the in-plane expansion
stalls, while the expansion in the transverse direction can
develop.  On the other hand, the~expansion is slow,
the~spectator pieces do not present much inhibition and then
expansion can continue primarily in the in-plane direction
untill the complete decompression.

The~presence of the spectator pieces allows then to assess the
pace of
expansion of the participant matter from the elliptic
flow for participants.  Figure~\ref{fig:fig3}
shows the elliptic flow excitation functions for different
calculations.  The flow turns more in-plane as the beam energy
increases and the spectator pieces move faster away from the
reaction zone and get shortened by the Lorentz contraction.
For
the cascade calculation, for which the expansion is sluggish,
the~flow is of a~primarily in-plane character at the
displayed energies.  The~harder the EOS, the~more out-of-plane
the flow gets.  However, there is more sensitivity to the EOS
at
the low energies when the spectator pieces remain still close
to
the participant zone than at high.  In addition to the~EOS, the
momentum dependence of MFs felt by the emerging baryons
affects their flow.  The~momentum dependence makes the fields
felt by high-momentum particles more repulsive, speeding up
the expulsion of these particles from the central participant
zone.

\begin{figure}
\begin{minipage}[t]{80mm}
\centerline{\includegraphics[angle=0,origin=c,
width=79mm]{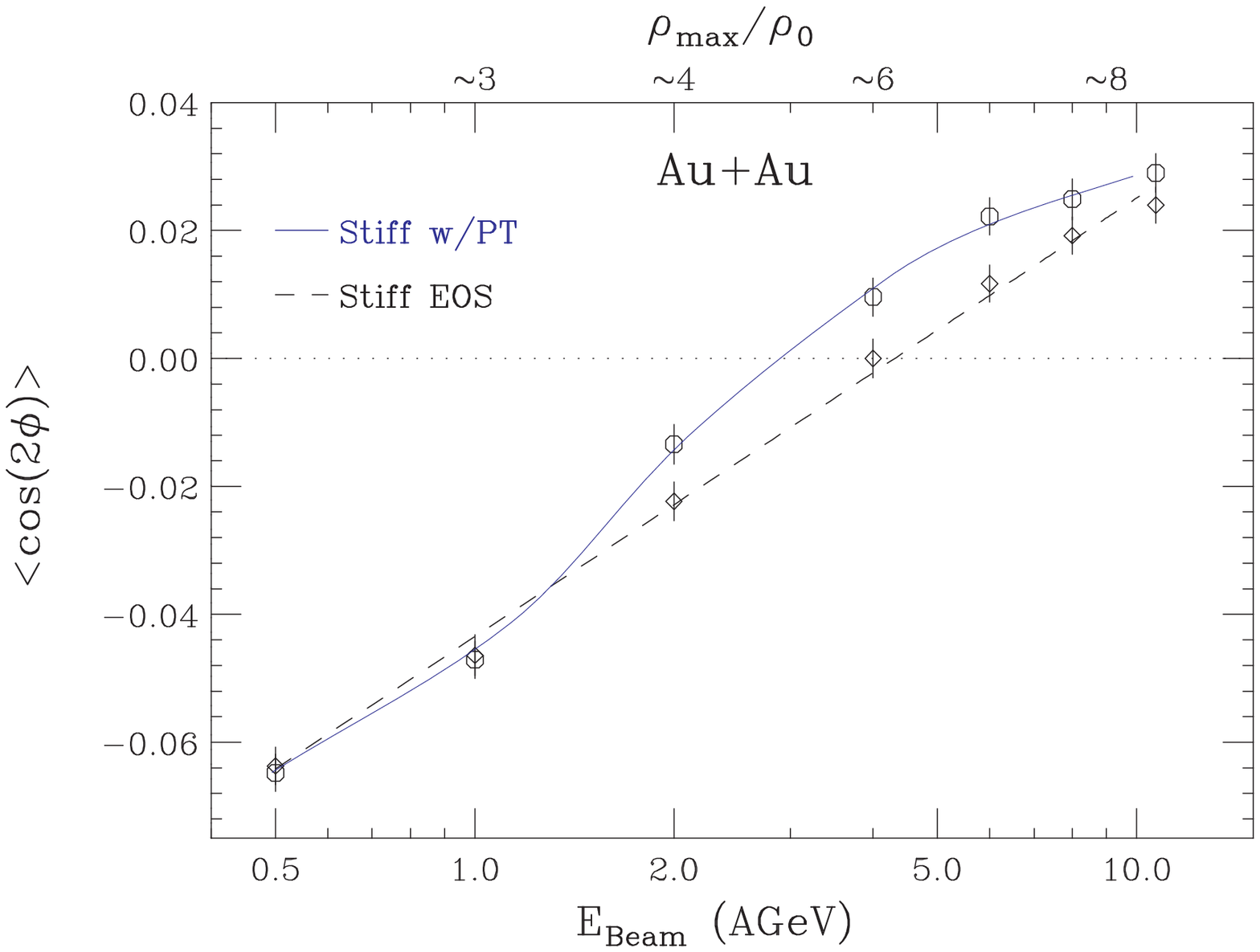}}
\caption{
                Calculated elliptic flow excitation functions for Au +
Au. The~diamonds represent results obtained with a~stiff EOS.
The~circles represent results obtained with a~stiff EOS
and with a~second-order
phase transition.  The~lines
guide the eye.
}
\label{figroy4a}
\end{minipage}
\hspace{\fill}
\begin{minipage}[t]{74mm}
\centerline{\includegraphics[angle=0,origin=c,
width=73mm]{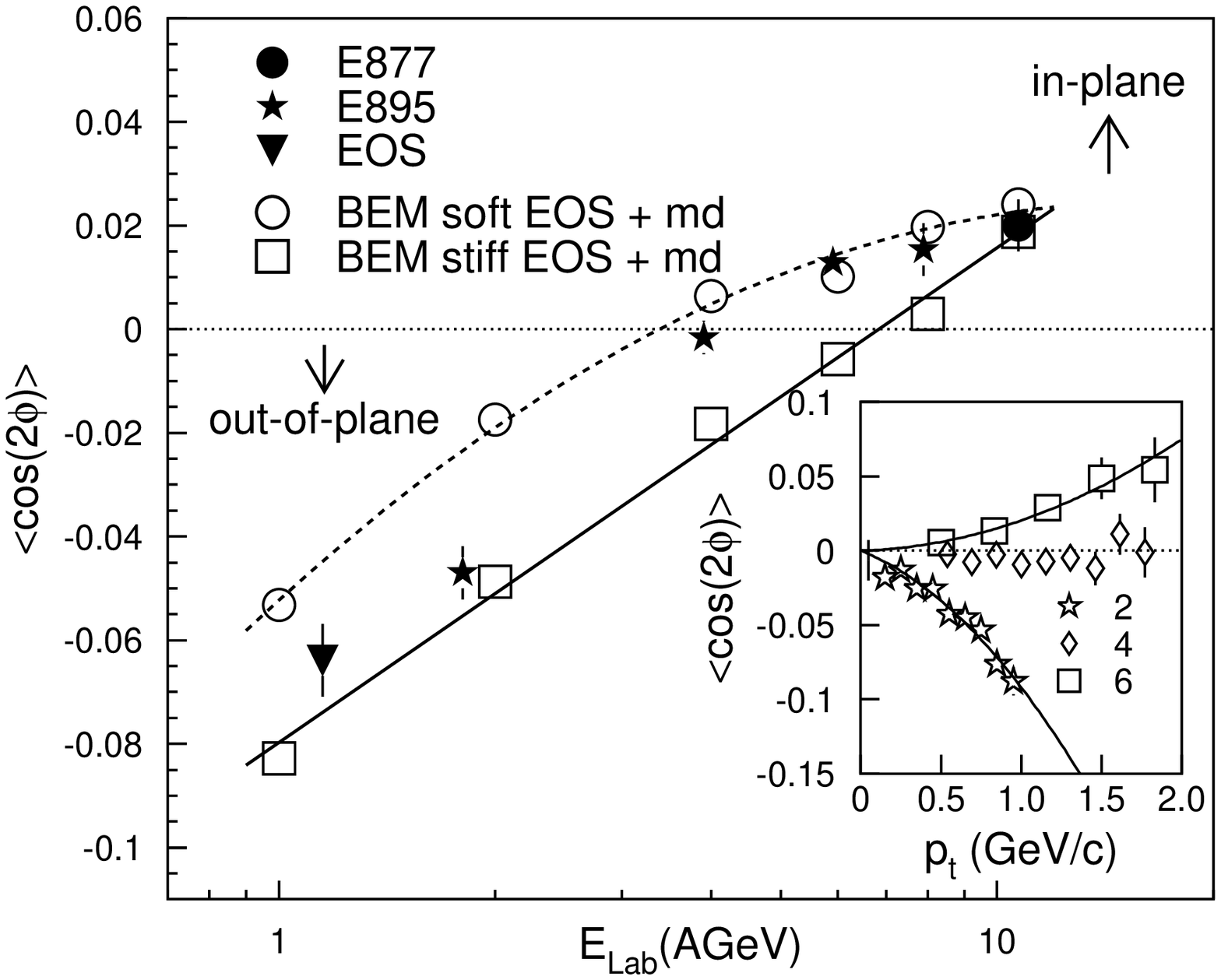}}
\caption{Elliptic-flow excitation function for Au + Au.  The
data
are from Refs.~\protect\cite{pin99}, \protect\cite{par99}, and
\protect\cite{bra98}.  Calculations have been carried out for
the soft and the stiff EOS with momentum
dependence.}
\label{fig:royel}
\end{minipage}
\end{figure}

Next question is whether a~phase transition could
sufficiently change the expansion as a~function of energy to
yield
an observable trace in the excitation function for ellipticity.
The answer is provided in Fig.~\ref{figroy4a}
that shows the excitation function for a~hard EOS with a
phase
transition which softens that EOS at higher densities similarly
to the case in Fig.~\ref{fig:eosnn00}.  At lower energies the
excitation function in Fig.~\ref{figroy4a} follows the hard EOS
function, but then it
switches to the soft EOS function as the densities reached in
the reactions increase.

Now the question is what kind of expansion scenario is followed
in Nature.  Comparing the data in Fig.~\ref{v2} to the cascade
calculations in Fig.~\ref{fig:fig3} at 1-4~GeV/nucleon, it is
apparent
that the expansion is much more rapid than in the cascade model
at the reached densities.
Figure~\ref{fig:royel} shows next a collection of the
data with momentum-dependent calculations superimposed.  It is
seen that at 2~GeV/nucleon the expansion is as rapid as for the
hard EOS with the momentum dependence, represented in
Fig.~\ref{fig:con}.  At higher energies, though, the~expansion
becomes more such as for the soft EOS with behavior as
a~function of energy strongly reminding that in
Fig.~\ref{figroy4a} for the phase transition.

\begin{figure} %[htb]
\centerline{\includegraphics[angle=90,origin=c,height=3.4in,
width=.70\linewidth]{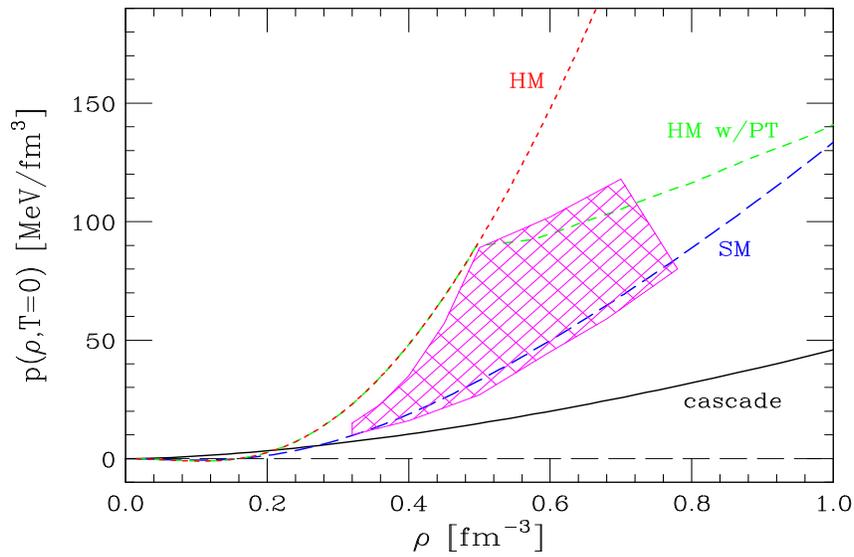}\vspace*{-.5in}}
\caption{Shadowed region shows boundaries on pressure vs baryon
density in $T=0$ symmetric nuclear matter, from flow
observables.}
\label{fig:eosdnn00}
\end{figure}

\begin{figure} %[htb]
\centerline{\includegraphics[angle=0,origin=c,height=3.3in,
width=.70\linewidth]{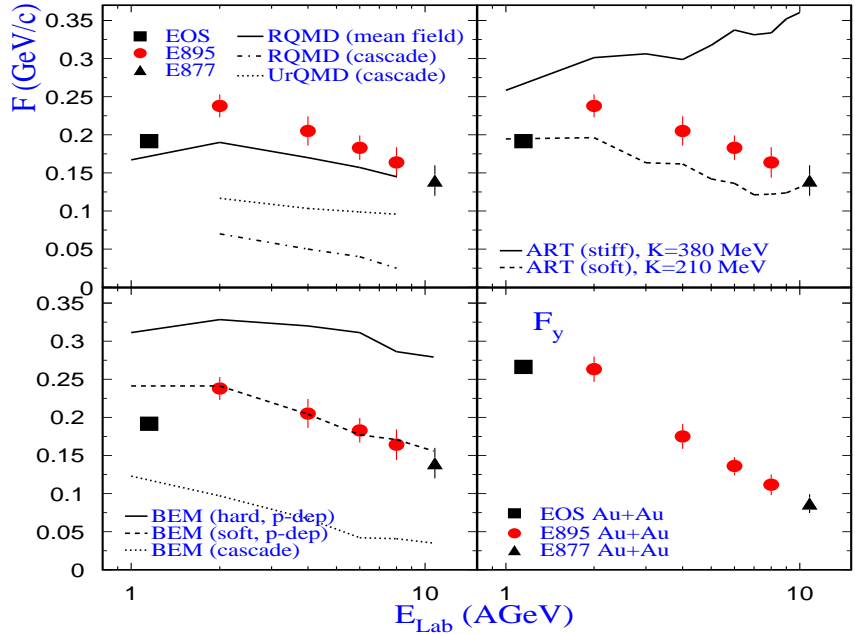}}
\caption{
Proton flow magnitude as a function of beam energy; the lower
right panel shows
the measured $F_y$, while the other three panels show
identical measurements of the parameter
$F$, with different transport model calculations
superimposed. The error bars include systematic uncertainties.
}
\label{fig:liu}
\end{figure}

Below 1~GeV/nucleon different flow data favor a~soft EOS with
momentum dependence, e.g.~\cite{pan93}.  Sideward data favor
that EOS in fact up to 10~GeV/nucleon; admittedly, though, the
sideward
flow is more exposed to the uncertainties in stopping than the
elliptic flow.  Combining older and the newer higher-energy
elliptic
and sideward flow results, we can draw broad boundaries for the
dependence of pressure on baryon density (or on energy density)
as exibited in Fig.~\ref{fig:eosdnn00}.
The~pressure is certainly higher than in the cascade
model; other than in the 1-2~GeV/nucleon region (vicinity of
the upper boundary kink in Fig.~\ref{fig:eosdnn00}) there is no
indication of any preference for the hard~EOS.
  A~set
of sideward flow data with model predictions superimposed is
shown in Fig.~\ref{fig:liu}.
Contrary to Fig.~\ref{fig:royel},
no evidence for a sudden softening of EOS is seen here.
As to conclusions on pressure vs baryon density, those
drawn for asymmetric matter from neutron star properties
generally pertain~\cite{lat00} to lower densities than in
Fig.~\ref{fig:eosdnn00}.

Deducing the properties of equilibrium matter requires
restricting the range of the dependence of baryonic MFs
on momentum.  This can be done by examining the dependence of
elliptic flow on momentum at the energies of interest.

\section{DIFFERENTIAL STUDIES OF FLOW}

\begin{figure}%[htb]
\begin{minipage}[t]{77mm}
\centerline{\includegraphics[angle=0,
width=76mm]{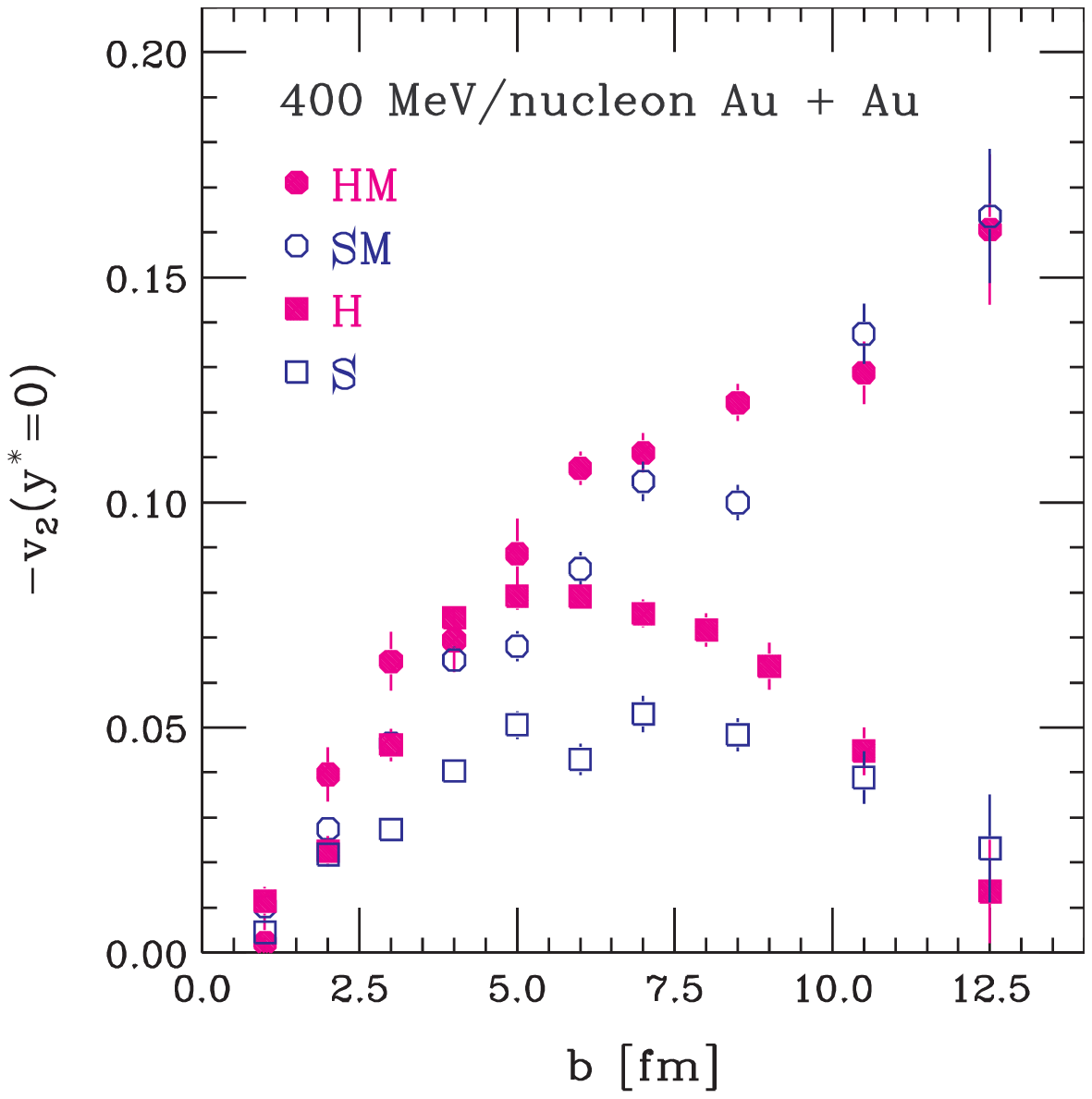}}
\caption{
Negative of the ellipticity coefficient
at midrapidity,
as a~function of the impact parameter,
from simulations of Au + Au collisions at 400~MeV/nucleon.
The~squares and circles represent, respectively, the~results
for MFs without and with momentum dependence.
The~filled
symbols
are for the incompressibility~$K=380$~MeV and the open symbols
are for~$K=210$~MeV.
}
\label{v2b}
\end{minipage}
\hspace{\fill}
\begin{minipage}[t]{77mm}
\centerline{\includegraphics[angle=90,origin=c,height=76mm,
width=76mm]{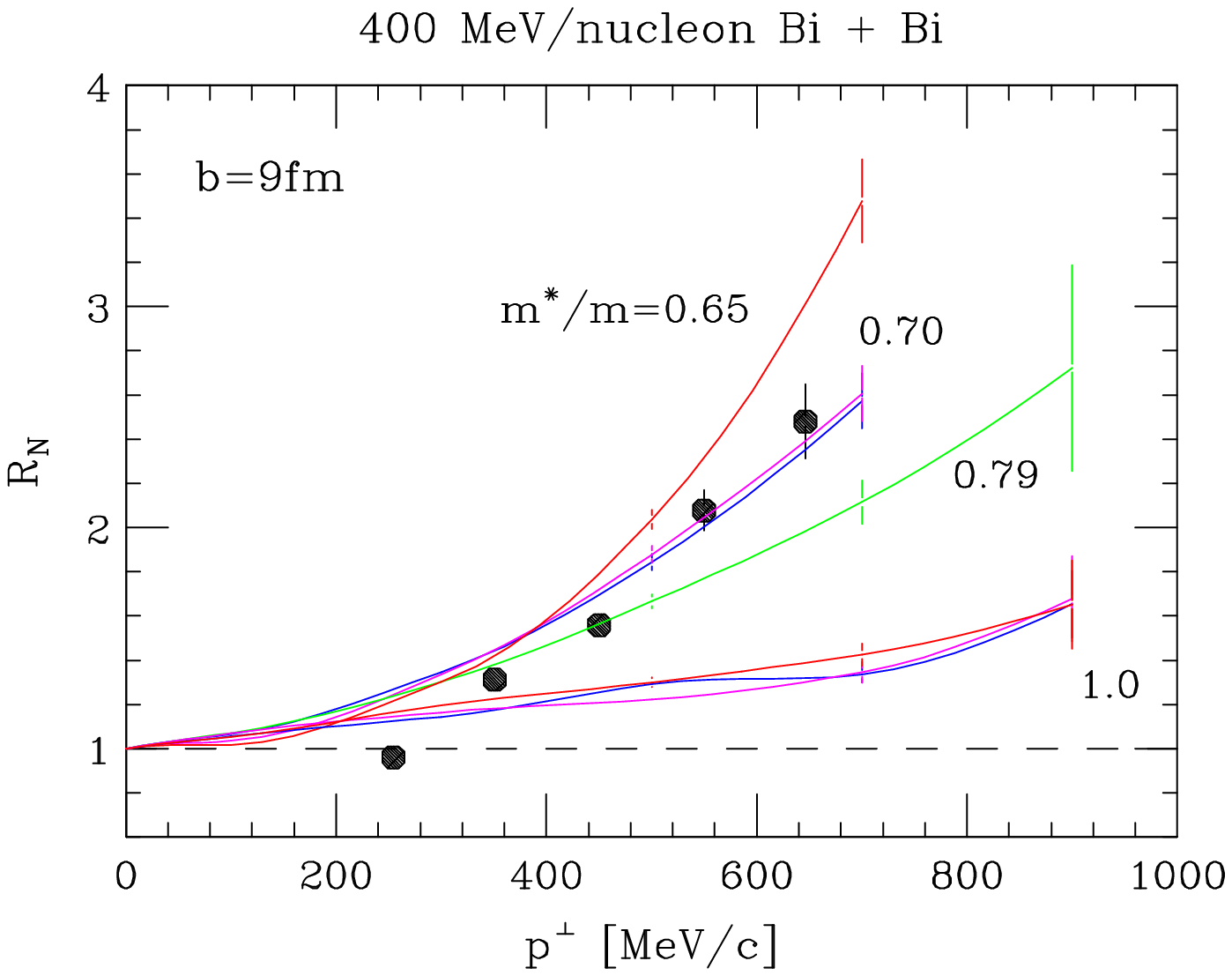}}
\caption{Out-of-plane to in-plane ratio $R_N = (1 - 2 \,
v_2)/(1
+ 2 \, v_2)$ for protons emitted at midrapidity from 400
MeV/nucleon Bi
+ Bi reactions, as a~function of transverse momentum.  Symbols
represent data of Ref.~\protect\cite{bri96} while lines
represent transport-model calculations for different
momentum dependencies in the MF.}
\label{RN}
\end{minipage}
\end{figure}

Studies of flow in narrow intervals of transverse momentum,
rapidity and impact parameters can help to disentagle the
effects of EOS and other characteristics of strongly
interacting matter.  Figure~\ref{v2b}
shows the dependence of $v_2$ on impact parameter in Au + Au
collisions at 400~MeV/nucleon.  At lower impact parameters,
the~effects of the stiffness of EOS and of mean-field momentum
dependence compete with each other.  However, at higher impact
parameters the predictions for the momentum-independent and
momentum-dependent MFs well separate.  Additional
sensitivity to the momentum dependence might be expected while
concentrating on flow at high momenta~\cite{pan93}.

Figure \ref{RN}
compares the ratio of out-of-plane to in-plane proton yields
\bed
R_N = {N(90^\circ) + N(270^\circ) \over N(0^\circ) +
N(180^\circ)} = {1 - 2 v_2 \over 1 + 2 v_2} \,
\eed
from KaoS measurements~\cite{bri96} to the results of
simulations.  It is apparent that the data permit tuning the MF
momentum dependence as labeled by the effective mass.

\begin{figure}
\centerline{\includegraphics[angle=0,height=4in,
width=.60\linewidth]{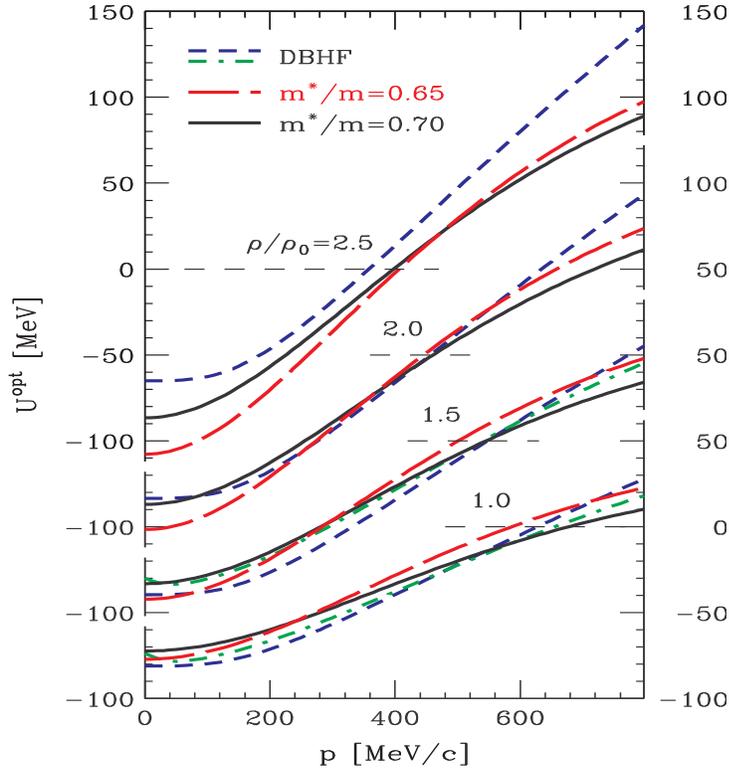}}
\caption{
Optical potential in nuclear matter as
a~function of nucleon momentum, at different densities, from
the DBHF
calculations with
the Bonn-A interaction~\cite{bro90} (short-dashed and
short-dash-dotted lines)
and in the transport-model parametrizations (solid and
long-dashed lines).
The~thin horizontal dashed lines indicate the zero
value for the potential.
}
\label{ugqli}
\end{figure}

Figure~\ref{ugqli}
compares next the transport-model parametrizations of MF
to the microscopic Dirac-Brueckner-Hartree-Fock (DBHF)
calculations with Bonn-A interaction.  Within the tested region
of densities there is a good agreement between the
parametrizations that best describe data~\cite{bri96} and
the microscopic calculations.  Other microscopic theories
exhibit a~variable degree of agreement with the optimal
parametrizations.

\section{CONCLUSIONS}

Flow observables are important tools for investigating
properties of hadronic matter in energetic collisions.
Measurements of second order and first order flow at beam
energies $\gtrsim 2$~GeV/nucleon show that the matter is
significantly stiffer than in the cascade model.  Lower and
also upper bounds on the pressure may be deduced as a function
of density.  Dependence of the elliptic flow on momentum probes
the dependence of mean field on momentum at supranormal
densities and may be used to constrain microscopic theories.
Some level of softening may occur in the EOS at the energies
between 2 and
4~GeV/nucleon, corresponding to a~compression
of 3-4~$\rho_0$.  Further insights
could be gained by studying the $b$-dependence of $v_2$.

\section*{ACKNOWLEDGEMENT}

This
work was partially supported by the National Science Foundation
under Grant PHY-0070818.

\end{document}